\documentstyle [12pt,epsfig]{article}
\begin{document}
\title{Estimate of the Energy of Upgoing Muons with Multiple Coulomb Scattering}

\author{D. Bakari $^{*}$, Y. Becherini, M. Spurio\\
 (For the MACRO Collaboration) \\
Dipartimento di Fisica dell'Universit\'a di Bologna and INFN \\
Viale Berti Pichat 6/2 - 40137 Bologna (Italy)\\
$^{*}$ {\small also LPTP, University Mohamed I, B.P. 524 Oujda (Morocco)} }

\maketitle
\begin {abstract}\small
\footnote{Invited contribution to the NATO Advanced Research Workshop: {\it Cosmic Radiation: From Astronomy To Particle Physics}, 21-23 March 2001, Oujda (Morocco)} 
Upward throughgoing muons in underground detectors are induced by atmospheric $\nu_\mu$ with energies in the $1-10^4\ GeV$ range. The oscillations of atmospheric muon neutrinos should affect mainly those with energy smaller than $E_\nu \sim 50\ GeV$, or $E_\mu \le 10\ GeV$ . We analyzed the MACRO\footnote{For a complete list of the MACRO Collaboration see the F. Ronga contribution to these Proceedings} upward throughgoing muons data set, using the Multiple Coulomb Scattering to estimate the energy $E_\mu$ of muons crossing the detector. We analyzed the event distribution for six event subsamples with different values of $<\ell_\nu / E_\nu>$. The event distribution is in agreement with the hypothesis of neutrino oscillations with $\Delta m^2$ few times $10^{-3}\ eV^2$ and maximum mixing.
\end {abstract}

\section{Introduction}

Neutrino oscillations, in a two neutrino mixing scenario, are the most likely solution for the {\it atmospheric neutrino problem} \cite{this01}.  
Underground experiments unfold the mass difference $\Delta m^2 = {m^2}_{\nu_3} - {m^2}_{\nu_2} $ and the mixing angle $\theta$ from the measurement of the survival probability 
$$P(\nu_\mu \rightarrow \nu_\mu)= 1 - sin^2 2\theta \, 
sin^2 (1.27 \Delta m^2  \ell_\nu / E_\nu)$$ 
The survival probability is usually plotted versus the {\it measurable} quantities $\ell_\nu$ and/or $E_\nu$. $\ell_\nu$ is the neutrino path length, estimated from the measurement of the zenith angle $\Theta$ of the daughter lepton.
% {\it L} = 2R cos$\Theta$. 
The neutrino energy $E_\nu$ is estimated from the calorimetric measurement of the energy of all final state particles. 

At present, $E_\nu$ is measured with some accuracy for low energy events (fully and partially contained SK \cite{Sk98} and Soudan2 \cite{Soudan2} events); however in this case $\ell_\nu$ is poorly reconstructed. The {\it vice versa} happens, when in the final state only the high-energy muon is measured (high energy neutrinos interacting in the rock around the detector, and inducing up-throughgoing muons in MACRO \cite{Macro98}, SK \cite{sk99} and Baksan\cite{miche}). The goal of next generation of atmospheric $\nu$ experiments should be the measurement of both $\ell_\nu$ and $E_\nu$, with enough precision to observe the first minimum of $P(\nu_\mu \rightarrow \nu_\mu)$. 

In the case of the MACRO experiment, the measured oscillations parameters 
(best value: $\Delta m^2 = 2.5 \times 10^{-3} \ eV^2$ and $sin^2 2\theta = 1$) are deduced: from the deficit and the distortion of the zenith angle distribution of up throughgoing $\mu$ ($i.e.$ $P(\nu_\mu \rightarrow \nu_\mu) \ vs. \ cos\Theta$) \cite{Macro95,Macro98}; from the measured deficit of lower energy partially contained events\cite{lownu}.

In this paper we present the following approach: we identified topologies of upgoing muons induced from neutrinos with different energies and path length $\ell_\nu$, and we discuss the results in terms of $\nu_\mu$ oscillations. The energy $E_\mu$ of detected muons is estimated through the multiple Coulomb scattering; $E_\mu$ is correlated with the energy $E_\nu$ of the parent neutrino. 
The survival probability shows a dependence with the $< \ell_\nu/E_\nu >$ parameter, unexpected without neutrino oscillations, while it is in good agreement with the quoted oscillation parameters. 

%%%%%%%%%%%%%%%%%%%%%%%%%%%%%%%%%%%%%%%%%%%%%%%%%
\section{Up throughgoing muon energy separation}
%%%%%%%%%%%%%%%%%%%%%%%%%%%%%%%%%%%%%%%%%%%%%%%%%
Up throughgoing muons are induced both by (relatively) low energy atmospheric $\nu_\mu$, interacting near the detector, and higher energy neutrinos, interacting farther. 
%The minimum energy for the $\mu$ to cross the whole detector is $\sim 1 \ GeV$. In Fig. 1a) is presented the distribution of the energy of the detected neutrino-induced up throughgoing muons (from Monte Carlo).
Assuming the neutrino oscillations mechanism, with 
$\Delta m^2 = 2.5 \times 10^{-3} \ eV^2$ and $sin^2 2\theta = 1$, a reduced flux of muons with $E_\mu \le 10\ GeV$ is expected, as shown by the shadowed histogram of Fig. 1a.

%%%%%%%%%%%%%%%%%%%%%%%%%%%%%%%%
\begin{figure}
 \begin{center}
 \centerline{\epsfig{figure= 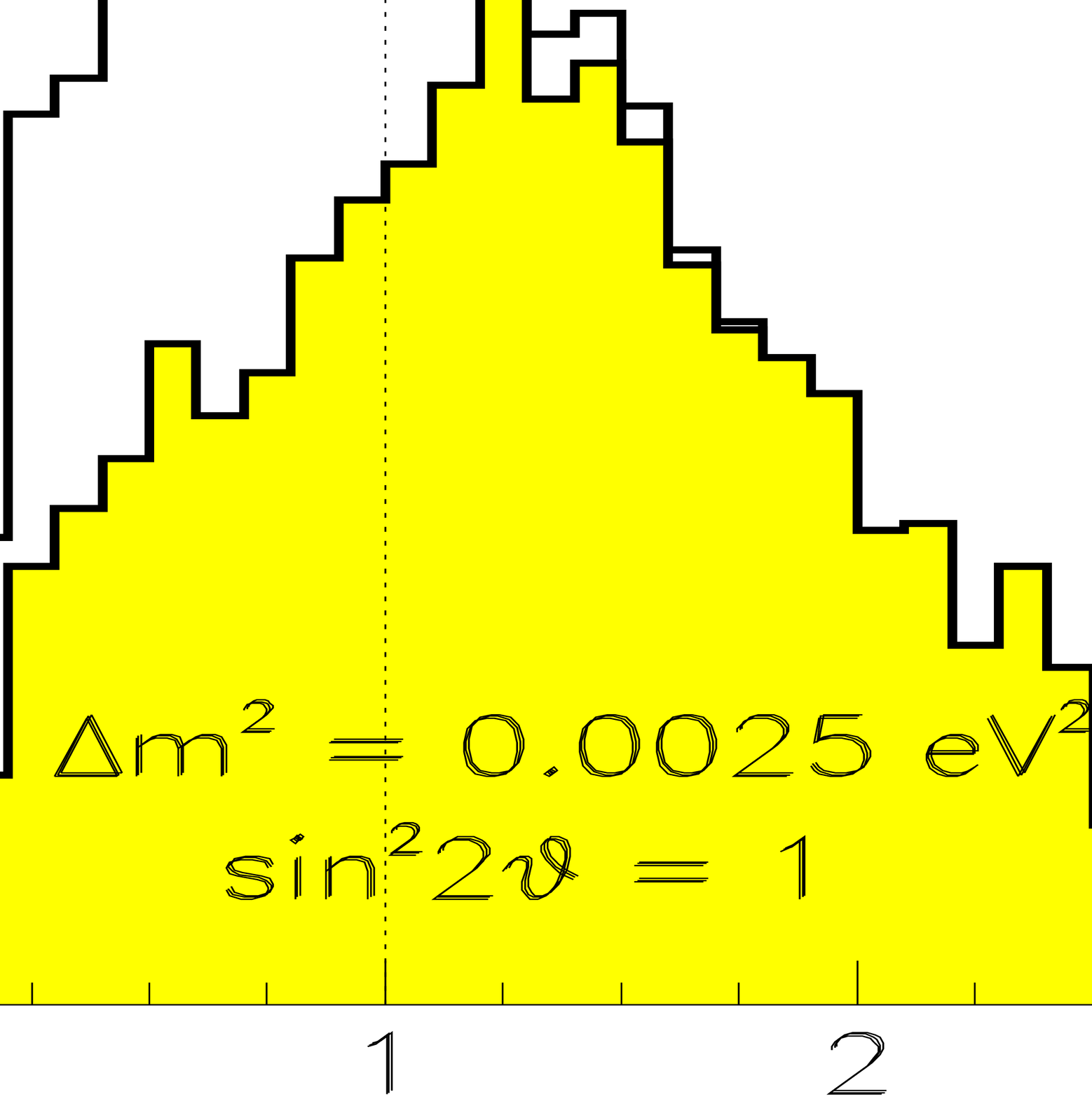,width=3cm,height=3cm} }
\vskip0.5cm
 \centerline{\epsfig{figure= 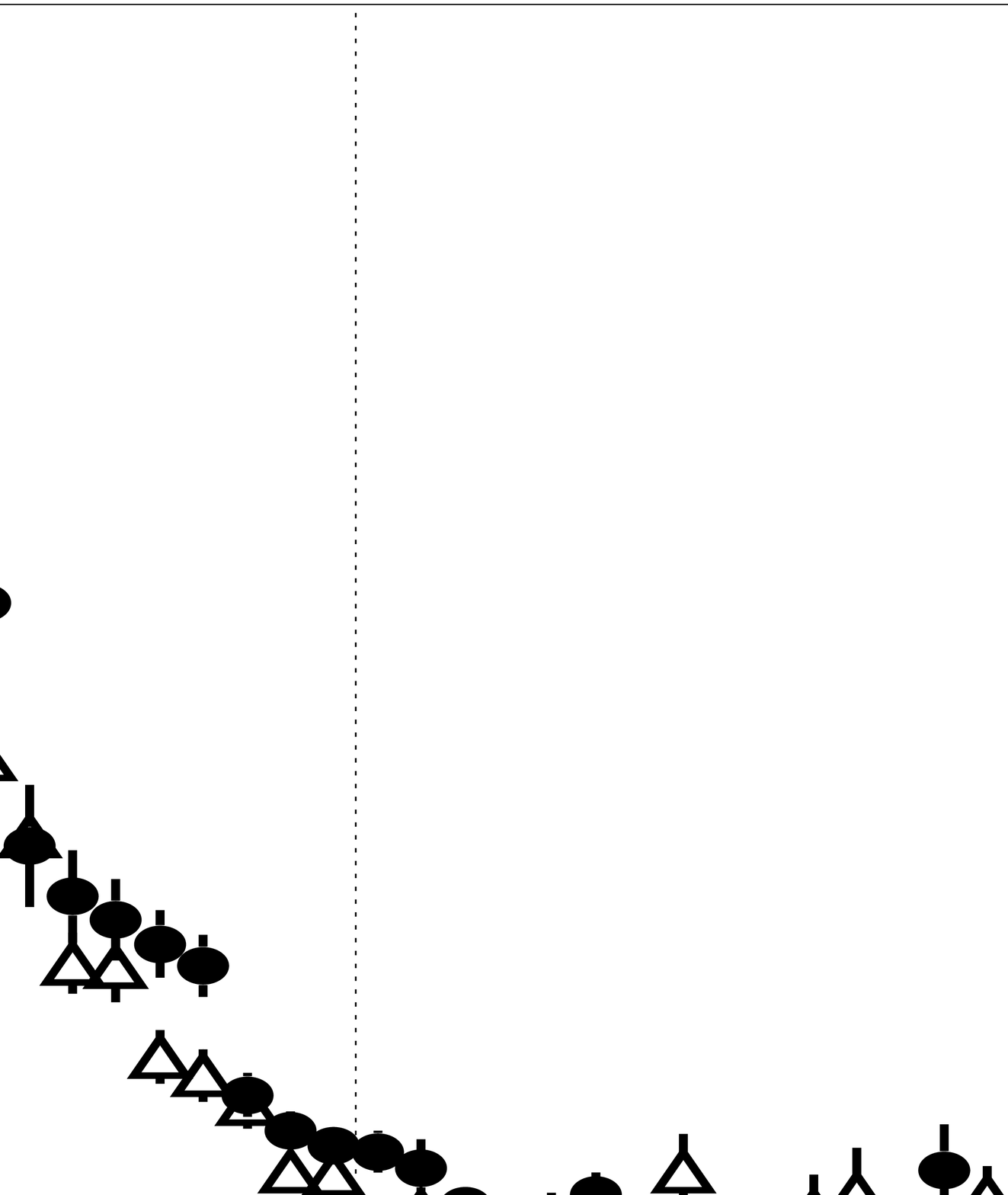,width=3cm,height=3cm} }
\end{center}
\caption{\small a) upper: energy spectrum of simulated neutrino-induced up throughgoing $\mu$ in MACRO. Full histogram for no oscillations; shadowed histogram, for $(\nu_\mu \rightarrow \nu_x)$ oscillations, with $\Delta m^2 = 2.5 \times 10^{-3} \ eV^2$ and $sin^2 2\theta = 1$.
b) lower: profile of the averaged value of the parameter $r_w\ vs.$ muon energy. The full (open) points refers to events generated with the $full$ ($table$) MC simulation, see text.}\vskip -0.5cm 
\end{figure}

\vskip 0.3 truecm
\noindent{\bf Multiple Coulomb Scattering}. Muons traversing the detector are deflected by many small-angle scatters, the bulk of which are Coulomb scatterings from the nuclei and from the atomic electrons (Multiple Coulomb Scattering, MCS). As an effect, muons are slightly deviated from a straight-line direction by an angular deflection (projected in a plane) $\phi_{plane}$ \cite{RPP}. 
% for muons crossing the lower part of MACRO,
Quantitatively, the averaged angular deflection $\phi^{rms}_{plane}$ and the spatial deflection $y^{rms}$ in one projective plane are approximated with: 
\begin{equation} \phi^{rms}_{plane} = {13.6\ MeV \over E_\mu} 
 \sqrt{x\over X_0} [1+0.04 ln({x \over X_0})]  \end{equation} 
\begin{equation} y^{rms} = 1/ \sqrt 3 \cdot  x \cdot \phi^{rms}_{plane} \end{equation}
In the case of MACRO, whose radiation length is $X_0 = 22.6\ gr/cm^2$ and total thickness $x =400\ gr/cm^2$, the average angular deflection is
$\phi^{rms}_{plane}  \simeq{0.052\ (rad) / E_\mu \ (GeV)}$. This corresponds to a projected spatial deflection in the wires view of the limited streamer tubes (whose cell size is 3 cm) of 
$ y^{rms} \simeq {7 (cm) / E_{\mu} (GeV)} $. {} {}
%\begin{equation} y^{rms} = 1/ \sqrt 3 \cdot  x \cdot \phi^{rms}_{plane} = 
%{7 (cm) \over E_{\mu} (GeV)}  \end{equation}
%The intrinsic spatial resolution of a MACRO streamer tube cell is 3 cm.
%, and the deviation of the muon sagitta from the straight line when the muon $crosses the whole detector can be measured, for $E_\mu$ up to $10 \ GeV$. 

\vskip 0.3 truecm
\noindent{\bf Experimental technique}. The technique we used to distinguish muons of low and high energy is illustrated in Fig. 2. All up throughgoing $\mu$, crossing the whole apparatus, were analyzed. 
The streamer tube hits in the lowest planes (N=1:8) were used for a refit, and the straight-line fit parameters were used to define the {\it lower track}. In a similar way, the highest planes (N=6:14) were used to define the {\it upper track}. The upper and lower tracks are used to evaluate $r_w$, the spatial difference in the $z=0$ plane between the two tracks, and $\Delta \Phi$, the angular difference between the slopes.
The value of the parameter $r_w$ ($\Delta \Phi$) depends on the muon energy  $E_\mu$.
In Fig. 1b the profile ({\it i.e.} the average value of $r_w$ evaluated in each bin of $E_\mu$ from our MC events) is shown. From the figure, the spatial difference between the lower and upper track exceed the intrinsic resolution of streamer tube cells when $E_\mu < 10\ GeV$. The behavior of the $\Delta \Phi$ parameter is analogous.

%%%%%%%%%%%%%%%%%%%%%%%%%%%%%%%%%%%%%%%%%%%%%%%%%
% SELEZIONE EVENTI + definizione cuts
%%%%%%%%%%%%%%%%%%%%%%%%%%%%%%%%%%%%%%%%%%%%%%%%%
\begin{figure}
%\vspace{5cm}  
\centerline{\epsfig{figure=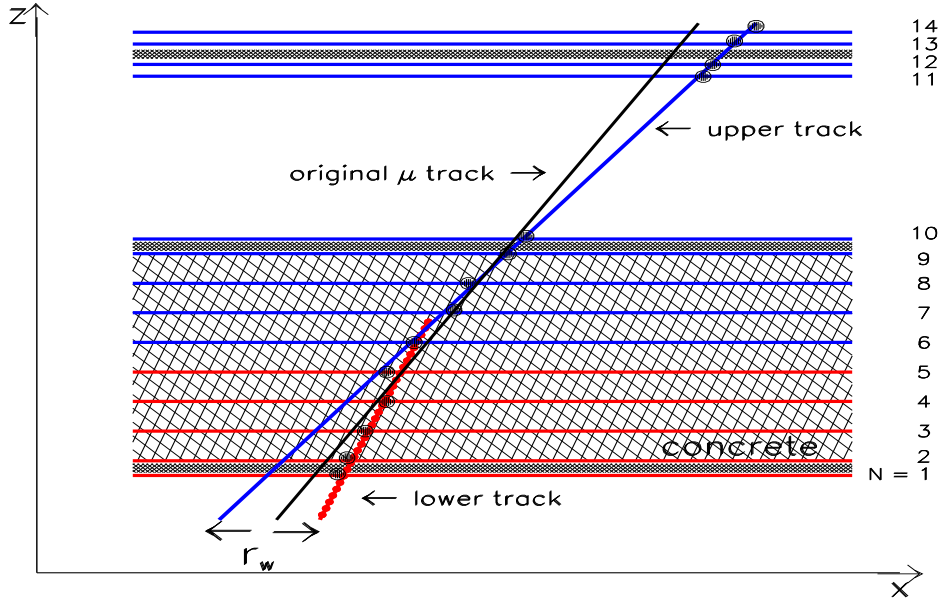,width=6cm,height=6cm} }
\caption{\small Sketch of the refit procedure to obtain a $lower$ and an $upper$ track, and the definition of the parameter $r_w$.}\vskip -0.5cm 
\end{figure}

\vskip 0.3 truecm
\noindent{\bf Event classification}.
A selection cut, based on $r_w$ and $\Delta \Phi$, was defined. 
Each up throughgoing muon is classified as:
{\it L= Low energy} if $r_w>3\ cm$ and $\Delta \Phi> 0.3^o$; 
{\it H= High energy} if  $r_w \le 3\ cm$ and $\Delta \Phi \le 0.3^o$.
The remaining events with ambiguous deviation, were classified as {\it M=Medium energy}. 
%In Fig. 2c is presented the distribution of energy for events classified as %{\it L} and {\it H} (and assuming no oscillations). 
Each sub sample has a different average energy for the parent neutrino. If $\nu_\mu$ disappearance is at work, the reduction should affect mainly the {\it L} sample, and we expect in the data a reduced $L\over H$ ratio, with respect to the MC prediction without oscillations.

\vskip 0.3 truecm
\noindent{\bf Checks on Monte Carlo predictions}.
A crucial point is the MC reliability, to avoid any systematic bias which can invalidates the results. For a detailed check of the detector simulation, we compared two large samples of measured with simulated downward going atmospheric muons.
Down throughgoing atmospheric muons have an average residual energy of $\sim 300\ GeV$ when reaching the underground laboratory, and 90\% of them have $E_\mu>10 \ GeV$. Our cut classified 63\% (25\%) of these events as $H$igh ($L$ow) energy. 
The small fraction of atmospheric muons ($\sim 0.3\%$), reaching MACRO with $E_\mu \le 1-5\ GeV$, stops inside the detector: 88\% (7\%) of them have been classified as $Low$ ($High$) energy. 

The energy distribution of these two atmospheric samples overlaps the spectrum of neutrino-induced upgoing $\mu$: the down-stop events in the lower, and the down- throughgoing in the high-energy tail. Using atmospheric muon samples, we adjusted the streamer tubes geometry in the MC database. After the MC correction, the measured and expected distribution of $r_w$ and $\Delta \Phi$ variables match together, both for the down-stop and down-throughgoing atmospheric muons. 
%This enables us to be confident that no systematic effects in the upgoing %samples can bias the data/MC comparison.

%%%%%%%%%%%%%%%%%%%%%%%%%%%%%%%%%%%%
\section{Results}
%%%%%%%%%%%%%%%%%%%%%%%%%%%%%%%%%%%%
\vskip 0.3 truecm
\noindent{\bf Real data.}
We used the same up throughgoing muon sample described in \cite{this01}a. The details of the events identification, cuts and selection efficiency were published \cite{Macro95,Macro98}. In our analysis, only the 316 events crossing the whole apparatus where used, to enhance the MCS effect on the muon track. The $r_w$ and $\Delta \Phi$ parameters are evaluated from the reconstructed {\it lower} and  {\it upper} tracks, and 101 events were classified by the cuts as {\it Low}, 51 as {\it Medium} and 164 as {\it High} energy.

\vskip 0.3 truecm
\noindent{\bf Simulated data.} The expected number of up throughgoing muons from $\nu_\mu$ interactions was evaluated with two different methods. In the first method ("table" MC) \cite{Macro95,Macro98}, the muon energy spectrum is obtained from a numerical convolution of the differential atmospheric neutrino spectrum, neutrino cross-sections and muon propagation in the rock. In the second ("full" MC), a full simulation is used: the $\nu_\mu$ interact in a large volume of rock around the detector, and the muon is propagated through the rock up to the detector.
%The same atmospheric neutrino flux (Bartol), DIS cross-section (GRVL094), muon %propagation in rock (Lohmann), and detector simulation in a GEANT-based code %are used in the two simulations. The first method is fast, but the neutrino %kinematics is lost. The second is very CPU consuming, but all information is %available. 
The simulated data have the same format of the real ones, and thus followed the same analyses steps. 
We applied the cut on $r_w$ and $\Delta \Phi$ on the selected up throughgoing MC muons: for a live time period equivalent of that of the real data, we expect 431 (430) events from "table" ("full") MC: 177 (178) {\it L}, and 191 (193) {\it H}.

\vskip 0.3 truecm
\noindent{\bf Uncertainties.} The overall theoretical uncertainty on MC predictions is 17\% [1,2]. This is a scale factor, which is almost constant when the number of events is plotted versus $cos\Theta$ or $E_\nu$. The shape of the distributions (as the angular one presented in \cite{Macro95,Macro98}) has a reduced theoretical uncertainty: we assumed a (conservative) value of 5\% in the considered range of $\nu$ energies.
The systematic uncertainties depend from many factor: the number of upper/bottom planes used in the refit; the choice of a particular value of $r_w$ and $\Delta \Phi$; the MC simulation statistics; the fluctuation in the streamer tube and scintillator efficiency along the running period; the detector acceptance uncertainty vs. the muon zenith angle. The evaluated overall systematic uncertainty is 7\%.

\vskip 0.3 truecm
\noindent{\bf Data and MC comparisons.}
We divided the events in three bins of $E_\mu$ $(L,M,H)$. The $L$ and $H$ samples were sub-divided in two bins of neutrino path length variable $\ell_\nu$ (the $M$edium energy sample have too few events). The number of detected and expected events $vs.$ $<log_{10}(\ell_\nu/E_\nu)>$ (MC evaluated) is presented in Table 1 and Fig. 3. It is also included an additional point, corresponding to the partially contained IU events \cite{lownu}. These events are induced by neutrinos with an average energy $E_\nu \sim 4\ GeV$, interacting inside the detector. This sample is contaminated by a 13\% contribution of $\nu_e$ and NC interactions.

\begin{figure}
\centerline{\epsfig{figure=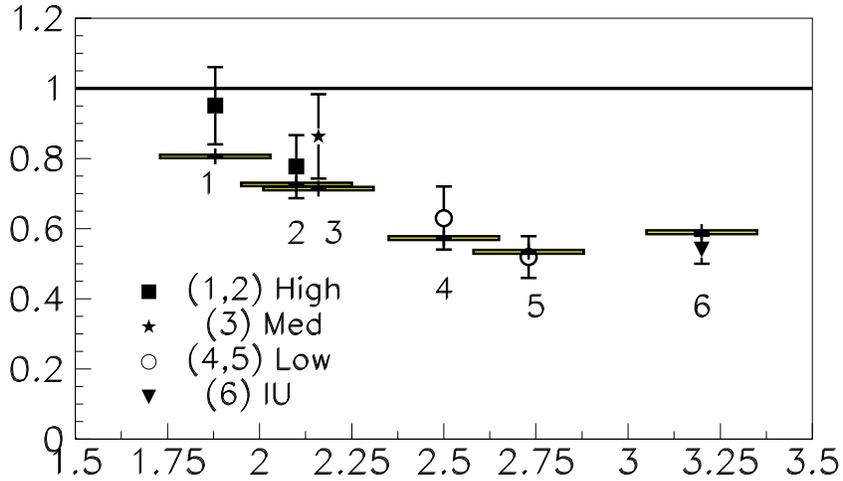,width=3.5cm,height=3.5cm} }
\caption{ \small Ratio R of the measured to the expected number of events vs. $<log_{10}(\ell_\nu/E_\nu)>$. The $H$ and $L$ samples are subdivided in two intervals of $\ell_\nu$. The black triangle is the point corresponding to the partially contained upgoing events is included [4]. The horizontal full line represents the expectation assuming no oscillations. The segment below each data point is the expectation assuming $\nu_\mu \rightarrow \nu_x$, with oscillation parameters quoted in Fig. 1a). }
\end{figure}

\begin{figure}
\centerline{\epsfig{figure=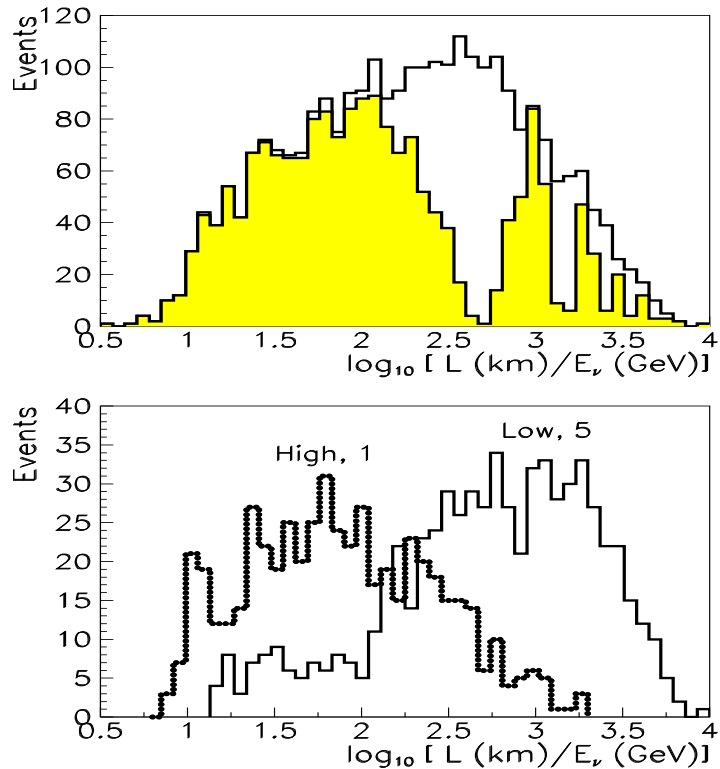,width=9cm,height=9cm} }
\caption{ \small 
a) Full histogram: simulated distribution of reconstructed up throughgoing muons in MACRO vs. $log_{10}(\ell_\nu/E_\nu)$. The shadowed region represents the expected distribution in case of neutrino oscillations (with the parameters of Fig. 1a). 
b) The full histogram corresponds to the distribution of simulated neutrino-induced muons, classified of high-energy from the Multiple Coulomb Scattering analysis, and with large zenith angle. The dashed histogram is for low-energy muons, with small zenith angle. The average value of these two distributions, are plotted in Fig. 3 (points 1 and 5).}
\end{figure}

\begin{table}[htb]
\begin{center}
\begin{tabular}{lcrrrr}
\hline
 & zenith     & $<log_{10}(\ell_\nu/E_\nu)>$ & $N_{data}$ & $N_{MC}$ & $N_{MC}^{osci}$  \\
\hline
H (1) & $|\cos\Theta|<0.8$      & $1.89\pm 0.55$ & $76$ & $80$  & $64$  \\
H (2) & $0.8<|\cos\Theta| <1.0$ & $2.11\pm 0.55$ & $88$ & $113$ & $82$  \\
M (3) & $0<|\cos\Theta|<1.0$    & $2.16\pm 0.66$ & $51$ & $59$  & $42$  \\
L (4) & $|\cos\Theta |<0.8$     & $2.50\pm 0.56$ & $47$ & $74$  & $43$  \\
L (5) & $0.8<|\cos\Theta|<1.0$  & $2.72\pm 0.60$ & $54$ & $104$ & $56$  \\
IU (6)& $0<|\cos\Theta|<1.0$    & $\sim 3.2$  & $154$ &$285$ & $168$  \\
\hline
\end{tabular}
\caption{\small Number of detected and expected events ($N_{data}$= detected, $N_{MC}$= expected and $N_{MC}^{osci}$=expected with oscillations). In column 3, there is the average value $\pm$ the HWHM of the $log_{10}(\ell_\nu/E_\nu)$ distribution.}
\end{center}
\end{table}

The ratio R between the detected to the expected number of events for each sub sample is shown in Fig. 3. The value of R expected with the used flux and neutrino cross sections, is 1 (no oscillations).  This value has an overall scale factor uncertainty of 17\%, which is almost constant all over the $(\ell_\nu/E_\nu)$ interval. 
In our data, R decreases when $<log_{10}(\ell_\nu/E_\nu)>$ increases. This behavior is consistent with the neutrino disappearance probability, when folded with our distribution of events. In Fig. 3 the segment below each data point is obtained from Monte Carlo simulated events, assuming oscillation parameters $\Delta m^2 = 2.5 \times 10^{-3} \ eV^2$, $sin^2 2\theta = 1$. 
The effect of neutrino oscillations in the distribution of $(\ell_\nu/E_\nu)$ is shown in Fig. 4a as a shadowed histogram. The effect of our cuts on the simulated events is shown in Fig. 4b for two subsamples (the one with the lowest/highest average value of $(\ell_\nu/E_\nu)$). 

We evaluated the $\chi^2$ for the shape of the ratio R vs. $<log_{10}(\ell_\nu/E_\nu)>$ for the two different test hypotheses: T0= no and T1= neutrino oscillations with the above quoted parameters. For hypothesis T0, we expect a (almost) constant value. Normalizing the number of expected with the number of detected events (316/430), we get a $\chi^2/dof = 14.8/5$ (probability of less than 2\%). For hypothesis T1 we expect a value of R which decreases as a "ladder" from 1 to 1/2, when $(\ell_\nu/E_\nu) >1000\ Km/GeV$. In this case, we get $\chi^2/dof = 4.7/5$ (probability=45\%) 

Hypotheses T0 and T1 can be tested also using the ratio of ratios $ ({L\over H})_{data}/ ({L\over H})_{MC}$. With the double ratio, most of the theoretical and systematic uncertainties cancel. For T1, we expect a larger reduction of events with $L$ow energy with respect to the ones with $H$igh energy. 
From our data, $ {({L\over H})}_{data}= 0.62\pm 0.08$. For T0, $({L\over H})^{T0}_{MC}= 0.92\pm 0.08$ (c.l. of 0.5\%). For T1 $({L\over H})^{T1}_{MC}= 0.68\pm 0.06$ (c.l. \ 90\%).

%%%%%%%%%%%%%%%%%%%%%%%%%%%%%%%%%%%%
\section{Conclusions}
We discussed the possibility to study the present high-energy atmospheric neutrinos data in terms of different bins of $<log_{10}(\ell_\nu/E_\nu)>$ , using the Multiple Coulomb Scattering on muons. The MACRO events were analyzed:  the result is consistent with the hypothesis of neutrino disappearance with $\Delta m^2 = 2.5 \times 10^{-3} \ eV^2$ and $sin^2 2\theta = 1$. This strengthens the hypothesis of neutrino oscillations to explain the data, against alternative more exotic models. A global analysis of all MACRO neutrino-induced data sets (which includes the results of this work) is in progress.

\vskip -0.2cm 
%%%%%%%%%%%%%%%%%%%%%%%%%%%%%%%%%%%%

\end{document}